\documentclass{article}

\usepackage{graphicx}%
\usepackage{multirow}%
\usepackage{amsmath,amssymb,amsfonts}%
\usepackage{amsthm}%
\usepackage{mathrsfs}%
\usepackage[title]{appendix}%
\usepackage{xcolor}%
\usepackage{textcomp}%
\usepackage{manyfoot}%
\usepackage{booktabs}%
\usepackage{algorithm}%
\usepackage{algorithmicx}%
\usepackage{algpseudocode}%
\usepackage{listings}%
\usepackage{setspace}
\usepackage{cleveref}
\usepackage{caption}
\usepackage{subcaption}
\usepackage{url}
\usepackage{natbib}
\usepackage{geometry}[margins = 2cm]

\newcommand{\by}{\boldsymbol{y}}
\newcommand{\bxi}{\boldsymbol{\xi}}
\newcommand{\blambda}{\boldsymbol{\lambda}}

\newcommand{\bE}{\boldsymbol{E}}
\newcommand{\be}{\boldsymbol{e}}
\newcommand{\bC}{\boldsymbol{C}}
\newcommand{\bY}{\boldsymbol{Y}}
\newcommand{\bV}{\boldsymbol{V}}
\newcommand{\bnu}{\boldsymbol{\nu}}
\newcommand{\bT}{\boldsymbol{T}}
\newcommand{\bI}{\boldsymbol{I}}
\newcommand{\bW}{\boldsymbol{W}}
\newcommand{\bU}{\boldsymbol{U}}
\newcommand{\bL}{\boldsymbol{L}}
\newcommand{\bv}{\boldsymbol{v}}

\theoremstyle{thmstyleone}%
\newtheorem{theorem}{Theorem}
\theoremstyle{thmstyletwo}%
\newtheorem{remark}{Remark}%

\theoremstyle{thmstylethree}%
\newtheorem{definition}{Definition}%

\newtheorem{lemma}{Lemma}[section]
\newtheorem{corollary}[theorem]{Corollary}

\usepackage{authblk}

\title{A Reduced Basis Decomposition Approach to Efficient Data Collection in Pairwise Comparison Studies}

\author[1]{Jiahua Jiang}
\author[1]{Joseph Marsh}
\author[1]{Rowland G Seymour\thanks{Corresponding author: \texttt{r.g.seymour@bham.ac.uk}}}
\date{}
\affil[1]{School of Mathematics, University of Birmingham, Edgbaston, B5 7US, United Kingdom}

\begin{document}
\maketitle


\abstract{Comparative judgement studies elicit quality assessments of objects through pairwise comparisons, typically analysed using the Bradley-Terry model. A challenge in these studies is experimental design, specifically, determining the optimal pairs to compare to maximize statistical efficiency. Constructing static experimental designs for these studies requires spectral decomposition of a covariance matrix over pairs of pairs, which becomes computationally infeasible for studies with a large number of objects. We propose a scalable method based on reduced basis decomposition that bypasses explicit construction of this matrix, achieving computational savings of two to three orders of magnitude. We establish eigenvalue bounds guaranteeing approximation quality and characterise the rank structure of the design matrix. Simulations demonstrate speedup factors exceeding 100 for studies with 64 or more objects, with negligible approximation error. We apply the method to construct designs for a 452-region spatial study in under 7 minutes, which was not previously possible, and enable real-time design updates for classroom peer assessment, reducing computation time from 15 minutes to 15 seconds.}



\maketitle

\section{Introduction}\label{sec1}
Comparative judgement is a type of study that has been widely used across the social sciences and humanities. In its simplest form, study participants are shown objects and asked which object has a higher quality. From these pairwise comparisons, a statistical model can be used to estimate the quality of each object and construct a ranked ordering of the objects. For example, in education research, comparative judgement studies have been carried out with teachers comparing students' mathematics homework based on which showed a better understanding of fractions \citep{Jones2015}; in Linguistics, language learners have been asked which of a pair of words in a foreign language was more difficult to learn \citep{Bisson2022}; and in one Philosophy study, participants were shown pairs of vehicles and asked which was worse a violation of a sign that reads ``no vehicles in the park'', in order to understand rule-violations \citep{Tanswell23}. 

The Bradley--Terry model \citep{Brad52} is perhaps the most commonly used statistical model to estimate the qualities of the objects in a comparative judgement study. This model assigns a quality parameter to each object in the study and estimates the value of these parameters from the pairwise comparisons. One limitation of the Bradley--Terry model is the difficulty of generating an experimental design. The main experimental design element in a comparative judgement study is deciding which pairs of objects to ask participants about. Experimental designs can take the form of scheduling distributions, which are a distribution over all possible pairs of objects, and determine the probability each pair of objects is shown to the study participant. Constructing the scheduling distribution for comparative judgement studies requires us to model the dependencies between comparisons, necessitating the consideration of not just pairs of objects, but \textit{pairs of pairs} of objects. This means the associated matrices are prohibitively large, dense, and exhibit a non-standard structure. Indeed, many studies are simply too large for current experimental design methods. A recent meta-analysis of 41 comparative judgement studies reported that 40\% contained more than 100 objects, while some citizen science studies have involved several thousand objects \citep{Kinnear2025}.

Well-designed experiments can substantially reduce the amount of data required, making comparative judgement studies more feasible in practice. Experimental designs for the Bradley--Terry models broadly fall into two categories: online designs (known as adaptive comparative judgement) and static designs that are constructed before the study begins. \citet{Pollitt2012} created an online experimental design, known as adaptive comparative judgement, where the objects featured in upcoming comparisons are determined based on comparisons made earlier in the study. Adaptive designs have come under criticism for two reasons. The first is that comparisons towards the end of the study are typically more difficult to make, as the adaptive design prioritises pairs that are close in quality \citep{Jones2023}. This can introduce an unwanted study effect, where data collected later on in the study period requires more effort to collect than data earlier on in the study. The second is that adaptive comparative judgement artificially inflates the reliability of the resulting parameter estimates \citep{Bramley2018}. Additionally, a limitation of adaptive designs is that they are only based on the quality parameters, and do not take into account any prior information or covariates. 


We focus on studies where the design is static and must be pre-calculated to account for specific constraints. This includes scenarios involving prior information, such as spatial correlations between objects, as well as sequential designs where the results of one round inform the scheduling of the next. Examples of sequential designs include sporting contests and educational peer-assessment activities, where students evaluate work in distinct stages separated by other learning activities.

Several methods exist for creating static designs, but can only be used for small studies or where no prior information is taken into account. \cite{Grass08} were able to analytically derive a static experimental design, but this is limited to studies with at most three objects. More recently, \citet{Roettger2025} develop a method for constructing the D-optimal design, but not in a Bayesian framework. \cite{Seymour25} developed a static design for studies where the objects are \textit{a priori} correlated. This method creates a scheduling distribution that specifies the probability of each pair of objects being compared. However, constructing this distribution is computationally expensive and can only be computed for studies with fewer than 150 objects, and it may take several hours or days to construct the distribution. These limitations come about because we need to construct and decompose the large, dense matrix describing the correlation between pairs of pairs of objects.

We develop a method that overcomes these computational barriers, allowing us to construct the scheduling distribution without constructing or decomposing the pairs-of-pairs correlation matrix. This allows us to construct the scheduling distribution in seconds or minutes, rather than hours, and for studies with several hundred objects, or even larger. This makes creating experimental designs possible for a much wider range of social science studies, which either contain a large number of objects, or are constrained by time and computational resources. 

In Section \ref{sec: Bradley--Terry model} we describe the Bradley--Terry model and the experimental design framework. Then in Section \ref{sec: RBD}, we develop a reduced basis decomposition method for approximating the design. We prove a relationship between the true design and our approximation, bounding its error and establishing conditions when the error is 0. In Sections \ref{sec: simulation study} and \ref{sec: application}, we apply our method to simulated and real data sets respectively, demonstrating the negligible error and considerable efficiency it provides. Finally, in Section \ref{sec: discussion}, we discuss the advantages and limitations of our method, as well as how it could be used to construct Bayesian D-Optimal designs. 

\section{The Bradley--Terry model} \label{sec: Bradley--Terry model}
We now describe the Bradley--Terry model and an experimental design for comparative judgement studies. Consider a study with $N$ objects, where the quality of object $i$ is denoted $\lambda_i \in \mathbb{R}$, and large positive values of $\lambda_i$ correspond to high quality objects, and large negative values to low quality objects. Examples of this include educational studies, where the objects being compared are students' coursework and the qualities represent the students' understanding, or psycholinguistic studies, where the objects are words in a foreign language and the qualities represent how difficult they are to learn. 

\subsection{Likelihood function and inference}
Consider a comparison between objects $i$ and $j$. In the Bradley--Terry model, the probability that object $i$ is judged to have a higher quality than object $j$ is defined as 
$$
p_{ij} = \frac{e^{\lambda_i}}{e^{\lambda_i} + e^{\lambda_j}} \iff \textrm{logit}(p_{ij}) = \lambda_i - \lambda_j. 
$$
If object $i$ and $j$ are compared $n_{ij}$ times, then the number of times object $i$ is judged to have a higher quality than $j$ can be modelled using the Binomial distribution
$$
Y_{ij} \sim \textrm{Bin}(n_{ij}, p_{ij}). 
$$
Given the vector of observed comparisons $\boldsymbol{y}$ and the vector of object qualities $\blambda$, the likelihood function is given by
\begin{equation}
  f(\by\mid\blambda) = \prod_{i=1}^N\prod_{j=1}^{i-1} \begin{pmatrix} n_{ij} \\ y_{ij}
\end{pmatrix} p_{ij}^{y_{ij}} (1-p_{ij})^{n_{ij} - y_{ij}}.  \label{eq: BT likelihood}
\end{equation}
By Bayes' theorem, the posterior distribution is given by $\pi(\blambda \mid \by) \propto f(\by \mid \blambda)\pi(\blambda)$. To allow for prior correlation between the object quality parameters and to allow for an offline experimental design, we place a multivariate normal prior distribution on the quality parameters. This distribution is given by $\blambda \sim \rm MVN(\boldsymbol{\mu}, C)$. 

The structure of the covariance matrix $C$ is a modelling choice and should reflect \textit{a priori} assumptions about how the qualities of the objects are related. In some settings where data is collected in rounds, for example student's abilities in educational assessments or sporting teams quality in tournaments, the prior covariance matrix can be constructed from the results of previous rounds (e.g. via the Hessian of the maximum likelihood estimates, or the posterior covariance matrix). In contexts where there study items have covariate information, the covariates can be form the prior covariance structure, particularly if these are spatial covariates. Finally, in contexts where there is a natural prior ranking to the objects, this can be included in a Toeplitz type structure, where the covariance between items $i$ and $j$ depend on $|i-j|$.  

\subsection{Experimental design}

Static, or offline, experimental designs specify in advance which pairs of objects will be compared during a study. The design takes the form of a probability distribution over the set of all possible pairs, from which comparisons are sampled during data collection. We formalise this as follows.

\begin{definition}
A \emph{scheduling distribution} is a probability distribution $\mathcal{S}$ over the set of all possible pairs of objects $\{(i,j) : 1 \leq i < j \leq N\}$. The scheduling distribution assigns probability $q_{ij}$ to the pair $(i,j)$. During data collection, each comparison is drawn independently from $\mathcal{S}$, so that $q_{ij}$ represents the probability that objects $i$ and $j$ are presented to a judge for comparison.
\end{definition} 

The scheduling distribution determines how comparisons are allocated across pairs of objects. The simplest approach is a uniform scheduling distribution, where $q_{ij} = 2/[N(N-1)]$ for all pairs, treating all comparisons as equally informative. However, when prior information is available about the relationships between objects, some comparisons may be more informative than others. For example, if two objects are known a priori to have similar qualities, comparing them may yield little new information about their relative ranking. An efficient scheduling distribution concentrates probability mass on pairs that are expected to be most informative, thereby reducing the total number of comparisons required to achieve a given level of precision in the quality estimates.

The challenge in constructing an efficient scheduling distribution lies in the dimensionality of the problem. Since the Bradley-Terry model relies on pairwise differences in quality parameters, we must model the dependencies between comparisons. This requires considering not just pairs of objects, but pairs of pairs of objects, leading to covariance matrices of dimension $\frac{N(N-1)}{2} \times \frac{N(N-1)}{2}$. As $N$ grows, constructing and decomposing these matrices becomes computationally prohibitive.

In this paper, we focus on the design proposed in \citet{Seymour25}. This is a principal component based design, where the scheduling distribution places higher probability on pairs of objects that explain higher prior variance. This design has been shown to reduce the number of comparisons that need to be collected. Since the Bradley--Terry model relies on pairwise differences in quality parameters, we can derive the distribution of these differences. This is derived via an affine transformation of the prior distribution $\boldsymbol{\lambda} \sim \mathcal{N}(\boldsymbol{\mu}, C)$, yielding:
\begin{equation*}
    \boldsymbol{\lambda}_{\text{diff}} \sim \mathcal{N}(\boldsymbol{\nu}, \Delta).
\end{equation*}
Here, the vector $\boldsymbol{\lambda}_{\text{diff}} \in \mathbb{R}^{N(N-1)/2}$ consists of the quality differences for each pair of objects, such that the entry corresponding to pair $(i,j)$ is $\lambda_i - \lambda_j$ for $i < j$, where the condition $i<j$ is used solely to index unique pairs and does not imply an ordering of the quality parameters. The vector $\boldsymbol{\nu}$ represents the corresponding differences in means, and $\Delta \in \mathbb{R}^{\frac{N(N-1)}{2} \times \frac{N(N-1)}{2}}$ is the covariance matrix capturing the dependence structure between pairwise comparisons. The entries of $\Delta$, representing the covariance between two pairs $(i,j)$ and $(k,l)$, are given by:
\begin{align}
    \operatorname{Cov}(\lambda_i - \lambda_j, \lambda_k - \lambda_l) &= \operatorname{Cov}(\lambda_i , \lambda_k) - \operatorname{Cov}(\lambda_i , \lambda_l) \nonumber \\
    &\quad - \operatorname{Cov}(\lambda_j , \lambda_k) + \operatorname{Cov}(\lambda_j , \lambda_l).
    \label{eq:delta}
\end{align}

We perform a spectral decomposition of the covariance matrix such that $\Delta = U \Psi U^T$, where $\Psi$ is a diagonal matrix of eigenvalues and the columns of $U$ contain the corresponding eigenvectors. We order the eigenpairs $\{\psi^{(c)}, \mathbf{u}^{(c)}\}$ in descending order such that $\psi^{(c)} \geq \psi^{(c+1)} \geq 0$, ensuring non-negativity as $\Delta$ is positive semidefinite. The eigenvectors $\{\mathbf{u}^{(c)}\}$ form an orthogonal basis for $\mathbb{R}^{M}$, where each vector represents a principal component \citep{Mardia79}.

In this framework, the $c$-th principal component corresponds to the direction explaining the maximum remaining variance of $\boldsymbol{\lambda}_{\text{diff}}$ orthogonal to the preceding $c-1$ components. The proportion of the total variance explained by the $c$-th component is given by $\psi^{(c)} / \sum_d \psi^{(d)}$. Consequently, the term $(\mathbf{u}^{(c)}_r)^2$ represents the contribution of the $r$-th variable (a specific pairwise difference) to the $c$-th principal component.

We define  the probability that the pair $\{i, j\}$ is presented to a judge as:
\begin{align}
    q_{ij} = \frac{\sum_c \left(\mathbf{u}^{(c)}_{r}\right)^2 \psi^{(c)}}{\sum_d \psi^{(d)}},
\end{align}
where $r$ is the linear index of $\boldsymbol{\lambda}_{\text{diff}}$ corresponding to the pair $(i, j)$, defined as $r = \frac{N(N-1)}{2} - \frac{(N-i +1)(N-i)}{2} + j - i$ (assuming $i<j$). This probability is equivalent to the sum of the squared loadings for an object risk parameter weighted by the eigenvalues, normalized by the total variance. The resulting set $\{q_{1,2}, \ldots, q_{N-1,N}\}$ defines a probability distribution $\mathcal{S}$ over the set of all possible pairs, effectively assigning higher sampling mass to pairs exhibiting higher prior variance.

A workflow for the construction of $\mathcal{S}$ is shown in Algorithm \ref{alg:brute_force}. We note that this Algorithm only provides information about which comparisons should be collected and not how many comparisons need to be collected. This is still an area of active research in comparative judgement, however \citet{Kinnear2025} suggest that collecting ten times as many comparisons as there are items is required for reliable estimates for the quality parameters. 

\begin{algorithm}[ht!]
\caption{Construction of scheduling distribution $\mathcal{S}$}
\label{alg:brute_force}
\begin{algorithmic}[1]
\Require Objects $\{\lambda_i\}_{i=1}^N$, prior mean $\boldsymbol{\mu}$, prior covariance matrix $C$
\Ensure Scheduling distribution $\mathcal{S}$ over all possible pairings

\State Construct the distribution $\boldsymbol{\lambda}_{\text{diff}} \sim \mathcal{N}(\boldsymbol{\nu}, \Delta)$, where $\boldsymbol{\lambda}_{\text{diff}}, \boldsymbol{\mu} \in \mathbb{R}^{\tfrac{N(N-1)}{2}}$ and $C \in \mathbb{R}^{\tfrac{N(N-1)}{2} \times \tfrac{N(N-1)}{2}}$. 

\State Perform SVD on $\Delta$ such that $\Delta = U \Psi U^T$, where $\Psi = \operatorname{diag}\left(\psi_1, \ldots, \psi_{\tfrac{N(N-1)}{2}}\right)$.

\State Return eigenpair  $\{\psi_i, U(:,i)\}_{i=1}^{\tfrac{N(N-1)}{2}}$. 

\State Compute the design probabilities $q_{ij} = \frac{\sum_c \left(\mathbf{u}^{(c)}_{r}\right)^2 \psi^{(c)}}{\sum_d \psi^{(d)}}$. 
\end{algorithmic}
\end{algorithm}

\section{Reduced basis decomposition for the Bradley--Terry model} \label{sec: RBD}
Current static experimental designs become infeasible for large $N$ due to the quartic growth in memory and polynomial growth in compute time. Because $\Delta$ is a dense matrix of size $\frac{N(N-1)}{2} \times \frac{N(N-1)}{2}$, it imposes an $\mathcal{O}(N^4)$ storage overhead. This volume renders standard PCA impractical with its $\mathcal{O}(N^6)$ complexity \citep{jolliffe2011principal}. Furthermore, even truncated variants such as randomized PCA \citep{eckart1936approximation, halko2011finding}  and Lanczos algorithms \citep{lehoucq1998arpack} fail to sufficiently alleviate the cost, requiring $\mathcal{O}(N^4d)$ operations. To address this,  we propose a novel computational framework to efficiently construct $\Delta$ and compute its spectral decomposition by exploiting the reduced basis decomposition (RBD) method. Our approach has three main components:
\begin{enumerate}
    \item To circumvent the direct computation of eigenvalues for $\Delta$, our efficient construction method uses a large sparse matrix and the standard covariance matrix of $\{\lambda_i\}^N_{i=1}$ to represent $\Delta$. This reduces the storage cost of $\Delta$ from $\mathcal{O}(N^4)$ to $\mathcal{O}(N^2)$. Furthermore, this formulation not only exploits the low rank nature of $\Delta$ for large $N$, but also allows its rank to be derived directly from the standard covariance matrix (\cref{sec:thm}).
    \item Our RBD-based transformation maps the large sparse matrix to a low-dimensional representation, achieving an $\mathcal{O}(N^3d) (d \le N)$ complexity for the spectral decomposition of $\Delta$. This constitutes a significant acceleration, outperforming truncated PCA by one order of magnitude and standard PCA by two.
    \item As shown in our theoretical analysis, the approximations satisfy an interlacing relation with the eigenvalues of $\Delta$, ensuring a systematic approach toward the exact spectrum. Full accuracy is captured when the number of iterations reaches $N-1$.
\end{enumerate}
We summarize our approach in \cref{alg:rb_spectral} and elaborate on the individual steps in \cref{sec:con_delta}–\cref{sec:eigdelta}. The  theoretical analysis regarding accuracy, memory requirements, and computational costs is detailed in \cref{sec:thm}.

\begin{algorithm}[h!]
\setstretch{1.2}
\caption{Reduced-basis spectral decomposition of $\Delta$}
\label{alg:rb_spectral}
\begin{algorithmic}[1]
\Require Objects $\{\lambda_i\}_{i=1}^N$, target rank $d$
\Ensure Approximation of the largest $d$ eigenvalues and eigenvectors of $\Delta$

\State Efficiently construct $\Delta = \bE \bC \bE^\top$, where $\bC$ is the standard covariance matrix 
       $[\mathrm{Cov}(\lambda_i, \lambda_j)]_{i,j=1}^N$, and $\bE$ is a sparse matrix; see \cref{sec:con_delta}.

\State By RBD, $\bE$ can be approximated via $\widetilde{\bE} = \bY \bT$, 
       with $\bY \in \mathbb{R}^{\frac{N(N-1)}{2} \times d}$ containing orthonormal columns,
       $\bT \in \mathbb{R}^{d \times N}$; see \cref{sec:rbd}.


\State Perform the SVD of the projected matrix $\widetilde{\bC} = \bT\bC\bT^\top \in \mathbb{R}^{d \times d}$. Let the singular values of $\widetilde{\bC}$ be $\{\sigma_i\}_{i=1}^d$  and the singular vectors be the columns of $\bV \in \mathbb{R}^{d \times d}$. The largest $d$ eigenvalues  of  $\Delta$ can be approximated by $\{\sigma_i\}_{i=1}^d$, and the corresponding eigenvectors can be approximated by $\bY \bV$; see \cref{sec:eigdelta}.
\end{algorithmic}
\end{algorithm}

\subsection{Efficient construction of $\Delta$}\label{sec:con_delta}
Let $\be_i = (0, 0, \dots, 0, 1, 0, \dots, 0) \in \mathbb{R}^{1\times N}$ denote the i-th standard basis vector. Let $\bC$ be the covariance matrix of variables $\{\lambda_s\}_{i=1}^N$, where each entry is given by $ \bC_{sh} = \text{Cov}(\lambda_s, \lambda_h)$. By the definition of $\Delta$ in \cref{eq:delta}, we have
\begin{align}
    \text{Cov}(\lambda_i - \lambda_j, \lambda_k - \lambda_l) & = \be_i \bC \be^{\top}_k - \be_i \bC \be^{\top}_{\ell} - \be_j \bC \be^{\top}_k + \be_j \bC \be^{\top}_{\ell} \\
    & = (\be_i - \be_j)\bC (\be_k - \be_{\ell})^{\top},
\end{align}
 where $\be_k$ is the $k-$th standard basis vector and $1\le i <j \le N, 1\le k < \ell \le N$. Hence, $\Delta$ can be represented as the matrix product:
\begin{align}
\label{eq:deltadef} 
\Delta = \bE \bC \bE^\top
\end{align}
where $\bE \in \mathbb{R}^{\frac{N(N-1)}{2}\times N}$ is the matrix whose rows are the vectors $\be_i - \be_j$, indexed consistently with the entries of $\Delta$.

\subsection{Reduced-basis spectral decomposition of $\bE$}
\label{sec:rbd}
Note that $\bE$ is sparse and severely ill-posed, which motivates the use of dimension-reduction techniques to construct an accurate low-dimensional surrogate. Among such techniques, the RBD is a state-of-the-art approach that is substantially more efficient than PCA/SVD-based methods, particularly when $N$ is large. A further advantage of RBD is the availability of an a posteriori error estimator that quantifies the discrepancy between the reduced representation and the original high-dimensional data.

Let $d$ denote the number of iterations. Given a maximal reduced dimension $d_{\max}$ satisfying $d_{\max} \le \operatorname{rank}(\bE) \le N-1$ (see \cref{lem:rankE}), and a tolerance $\varepsilon_{R}$ governing the admissible approximation error, the RBD procedure returns a basis matrix $\bY \in \mathbb{R}^{\frac{N(N-1)}{2}\times d}$ spanning the reduced space, along with a coefficient matrix $\bT \in \mathbb{R}^{d\times N}$. Here, $d \le d_{\max}$ denotes the actual dimension of the compressed data. The approximation of $\bE$ is then given by
\begin{equation}
\label{eq:rbdE}
\widetilde{\bE} = \bY\bT.
\end{equation}
The basis matrix $\bY$ is constructed iteratively so that its column span provides increasingly accurate approximations to the column space of $\bE$. The procedure is initialized with either a randomly chosen column of $\bE$ or a user-specified vector. At iteration $k$, given the current basis ${\by_1,\dots,\by_k}$, the next vector $\by_{k+1}$ is chosen by scanning all columns of $\bE$ and selecting the one whose projection onto $\operatorname{span}{\by_1,\dots,\by_k}$ yields the largest residual. This enrichment continues until either the maximal projection (or compression) error falls below the prescribed tolerance $\varepsilon_R$ or the dimension reaches $d_{\max}$.   An additional computational benefit of RBD lies in its offline–online decomposition, which enables fast evaluation of the error estimator once appropriate quantities have been precomputed.
A complete description of the algorithm is given in \cref{alg:rbd}.


\begin{algorithm}[h!]
\setstretch{1.2}
\caption{Reduced Basis Decomposition $(\bY, \bT) = \mathrm{RBD}(\bE, \varepsilon_{R}, d_{\max})$}
\label{alg:rbd}
\begin{algorithmic}[1]
\State Set $d = 1$, $\bE_{\text{cur}} = +\infty$, and $i$ a random integer between $1$ and $n$.
\While{$d \leq d_{\max}$ and $\bE_{\text{cur}} > \varepsilon_{R}$}
    \State $\bnu = \bE(:, i)$
    \State Apply the modified Gram-Schmidt (MGS) orthonormalization:
    \For{$j = 1$ to $d - 1$}
        \State $\bnu = \bnu - (\bnu \cdot \bxi_j) \bxi_j$
    \EndFor
    \If{$\|\bnu\| < \varepsilon_{R}$}
        \State $\bY = \bY(:, 1 : d - 1)$
        \State $\bT = \bT(1 : d - 1, :)$
        \State \textbf{break}
    \Else
        \State $\bxi_d = \frac{\bnu}{\|\bnu\|}$
        \State $\bY(:, d) = \bxi_d$
        \State $\bT(d, :) = \bxi_d^\top \bE$
    \EndIf
    \State $\bE_{\text{cur}} = \max\limits_{j \in \{1,\dots,n\}} \|\bE(:, j) - \bY(:, 1:d) \bT(:, j)\|$
    \State $i = \arg\max\limits_{j \in \{1,\dots,n\}} \|\bE(:, j) - \bY(:, 1:d) \bT(:, j)\|$
    \If{$\bE_{\text{cur}} \leq \varepsilon_{R}$}
        \State $\bY = \bY(:, 1:d)$
        \State $\bT = \bT(1:d, :)$
    \Else
        \State $d = d + 1$
    \EndIf
\EndWhile
\end{algorithmic}
\end{algorithm}

The RBD method is based on a greedy algorithm commonly used in model-reduction techniques \citep{ngoc2005certified, chen2013reduced, rozza2008reduced}. It builds the reduced space dimension-by-dimension: At each step, the \textit{greedy} algorithm identifies the column of $\bE$ that is least well represented by the current reduced space and adds it as a new basis vector. This choice is guided by a computable error indicator that measures the discrepancy between each column of $\bE$ and its projection onto the current reduced space. By repeatedly selecting the direction with the largest error, the algorithm ensures that the reduced basis improves systematically and targets the components of $\bE$ that are most difficult to approximate. 

\subsection{Eigenpairs approximation of $\Delta$}
\label{sec:eigdelta}
Let $\widetilde{\Delta}$ be the approximation of $\Delta$ via reduced-basis spectral decomposition. Define 
\begin{align}
\label{eq:defC}
    \widetilde{\bC} = \bT \bC \bT^\top,
\end{align}
where $\bC$ is the covariance matrix of $\{\lambda_i\}_{i=1}^N$ and $\bT$ is the coefficient matrix in \cref{eq:rbdE}. Then
    \begin{align}
        \widetilde{\Delta} &= \widetilde{\bE} \bC \widetilde{\bE}^{\top}\\
        &= \bY\bT \bC \bT^{\top}\bY^{\top}\\
        & = \bY \widetilde{\bC} \bY^\top,
    \end{align}
where $\bY \in \mathbb{R}^{\frac{N(N-1)}{2} \times d}$ is the basis matrix with orthonormal columns and $\bT \in \mathbb{R}^{d \times N}$ is the coefficient matrix.  Let the SVD of $\widetilde{\bC} \in \mathbb{R}^{d \times d}$ be given by
\begin{align}
\label{eq:svdC}
    \widetilde{\bC} = \bV \sum \bV^{\top},
\end{align}
where $\sum = \operatorname{diag}(\sigma_i)_{i=1}^d$ containing eigevalues $\sigma_i$ and $\bV \in \mathbb{R}^{d \times d}$ is a unitary matrix of eigenvectors. Then
\begin{align}
\label{eq:svddelta}
     \widetilde{\Delta} &= \bY \bV \sum \bV^{\top} \bY^\top.
\end{align}
Since $\bV^{\top} \bY^\top\bY \bV  = \bI_d$ (where $\bI_d$ is the identity matrix), 
$\{\sigma_i\}_{i=1}^d$ are the largest $d$ eigenvalues of $\widetilde{\Delta}$ and the columns of $\bY\bV$ form the corresponding eigenvectors. The remaining $N-d$ eigenvalues of $\widetilde{\Delta}$ are zero. 

\subsection{ Theoretical analysis}
\label{sec:thm}
In this section, we investigate the theoretical properties of the RBD approximation for the eigenpairs of $\Delta$ in the Bradley–Terry model over $d$ iterations. First, we present a theorem that shows the relation between eigenvalues of $\Delta$ and their approximations, eigenvalues of $\widetilde{\Delta}$.
\begin{theorem}
    Let $\{\alpha_i\}_{i=1}^{\frac{N(N-1)}{2}}$ be the eigenvalues of $\Delta$. Let $\{\sigma_i\}_{i=1}^d$ be the largest $d$ eigenvalues of $\widetilde{\Delta}$, which approximate the largest $d$ eigenvalues of $\Delta$. Both of them are sorted in non-increasing order. Then
    \begin{align}
        \alpha_{i + N-d} \le \sigma_i \le \alpha_i
    \end{align}
for $i = 1, \dots,d.$
\end{theorem}
\begin{proof}
    After running $d$ iterations (where $d \le d_{\text{max}} \le N$) of the RBD method (as defined in \cref{eq:rbdE}), we obtain
\begin{equation}
    \widetilde{\bE} = \bY\bT,
\end{equation}
where $\bY \in \mathbb{R}^{\frac{N(N-1)}{2} \times d}$ has orthonormal columns and $\bT \in \mathbb{R}^{d \times N}$ is the coefficient matrix.
If we keep running RBD for total $N$ iterations, then $\bE$ can be represented as 
\begin{align}
\label{eq:E}
    \bE = \begin{bmatrix}
        \bY & \bW
    \end{bmatrix}\begin{bmatrix}
        \bT \\ \bU
    \end{bmatrix},
\end{align}
where $\bW \in \mathbb{R}^{\frac{N(N-1)}{2} \times (N-d)} $ has orthonormal columns and $\bU \in \mathbb{R}^{(N-d) \times N}$. By \cref{eq:deltadef}, \cref{eq:defC} and \cref{eq:E},
\begin{align}
    \Delta &= \begin{bmatrix}
        \bY & \bW
    \end{bmatrix}\begin{bmatrix}
        \bT \\ \bU
    \end{bmatrix} \bC \begin{bmatrix}
        \bT^\top, \bU^\top
    \end{bmatrix}\begin{bmatrix}
        \bY^\top \\ \bW^\top
    \end{bmatrix} \\
    & = \begin{bmatrix}
        \bY & \bW
    \end{bmatrix} \begin{bmatrix}
        \bT \bC \bT^\top &  \bT \bC \bU^\top \\
        \bU\bC\bT^\top  & \bU \bC \bU^\top
    \end{bmatrix} \begin{bmatrix}
        \bY^\top \\ \bW^\top
    \end{bmatrix} \\ 
    & = \begin{bmatrix}
        \bY & \bW
    \end{bmatrix} \begin{bmatrix}
        \widetilde{\bC} &  \bT \bC \bU^\top \\
        \bU\bC\bT^\top  & \bU \bC \bU^\top
    \end{bmatrix} \begin{bmatrix}
        \bY^\top \\ \bW^\top
    \end{bmatrix}.
\end{align}
By \cref{eq:svdC} and \cref{eq:svddelta}, $\{\sigma_i \}_{i=1}^d$ are the eigenvalues of $\widetilde{\bC}$ and the largest $d$ eigenvalues of $\widetilde{\Delta}$. 
By Cauchy's interlacing theorem \citep{bellman1997introduction, horn2012matrix, parlett1998symmetric}, 
\begin{align}
        \alpha_{i + N-d} \le \sigma_i \le \alpha_i
    \end{align}
for $i = 1, \dots, d$. 
\end{proof}

Next, we introduce a lemma to compute $\operatorname{rank}(\bE)$ and then establish the consequential relation between $\operatorname{rank}(\Delta)$ and $\operatorname{rank}(\bC)$.

\begin{lemma}
\label{lem:rankE}
    Let $\bE \in \mathbb{R}^{\frac{N(N-1)}{2}\times N}$ be the matrix in efficient construction of $\Delta$ in \cref{eq:deltadef}, defined as 
    \begin{align}
        \bE\left(\frac{(N-i +1)(N-i)}{2} + j - i, :\right) = \be_i - \be_j
    \end{align}
for $1\le i < j \le N$, where $\be_k \in \mathbb{R}^{1 \times N}$ denotes the $k$-th standard basis vector. Then $\operatorname{rank}(\bE) = N-1$, and $\operatorname{Null}(\bE) = \operatorname{span}(\{\mathbf{1}\})$ with $\mathbf{1}  = (1,\dots,1)^\top \in \mathbb{R}^{N \times 1}$.
\end{lemma}
\begin{proof}
    Let $\bv  = (v_1, \dots, v_N ) \in \operatorname{Null}(\bE)$. Then \(\bE\bv = \mathbf{0}\), and in particular each row of \(\bE\) yields
\begin{align*}
    (\be_i - \be_j)^\top \bv &= 0, \\
    v_i& = v_j
\end{align*}
for $i\neq j \in \{1, \dots, N\}$.
Thus, \(\operatorname{Null}(\bE)=\operatorname{span}\{\mathbf{1} \}\) and $\operatorname{rank}(\bE) = N - \dim(\operatorname{Null}(\bE)) = N-1$.
\end{proof}

\begin{theorem}
    Let $\mathbf{1}$ be the vector of all ones in $\mathbb{R}^{N \times 1}$. Let $\Delta$ and $\bC$ be defined as above. If $\mathbf{1} \in \operatorname{range}(\bC)$, then $\operatorname{rank}(\Delta) = \operatorname{rank}(\bC) - 1$.  Otherwise, $\operatorname{rank}(\Delta) = \operatorname{rank}(\bC)$.
\end{theorem}
\begin{proof}
Since covariance matrix $\bC$ is symmetric semi-positive definite, let $\bC = \bL\bL^\top$ be the Cholesky Decomposition of $\bC$, where $\bL \in \mathbb{R}^{N \times N}$ is a lower triangular matrix with non-negative diagonal entries. Then 
\begin{align}
\label{eq:CL}
\operatorname{range}(\bC) = \operatorname{range}(\bL)
\end{align}
By \cref{eq:deltadef},
\begin{align}
\label{eq:EL}
\Delta = \bE\bL(\bE\bL)^\top.
\end{align}
Thus, by \cref{eq:CL} and \cref{eq:EL},
\begin{align}
\operatorname{rank}(\Delta) &= \operatorname{rank}(\bE\bL)\\
&= \operatorname{dim}\Big(\operatorname{range}(\bL)\Big) - \operatorname{dim}\Big(\operatorname{Null}(\bE) \bigcap \operatorname{range}(\bL)\Big) \\
& = \operatorname{dim}\Big(\operatorname{range}(\bC)\Big) - \operatorname{dim}\Big(\operatorname{Null}(\bE) \bigcap \operatorname{range}(\bC)\Big).
\end{align}
If $\mathbf{1} \in \operatorname{range}(\bC)$, by \cref{lem:rankE}, $\operatorname{range}(\bC) \bigcap \operatorname{Null}(\bE) = \operatorname{span}\{\mathbf{1}\}$. Then
\begin{align}
\operatorname{rank}(\Delta) &= \operatorname{dim}\Big(\operatorname{range}(\bC)\Big) - \operatorname{dim}\Big(\operatorname{span}\{\mathbf{1}\}\Big) \\
& = \operatorname{rank}(\bC) - 1.
\end{align}
If $\mathbf{1} \notin \operatorname{range}(\bC)$, by \cref{lem:rankE}, $\operatorname{range}(\bC) \bigcap \operatorname{Null}(\bE) = \emptyset$. Then
\begin{align}
\operatorname{rank}(\Delta) &= \operatorname{rank}(\bC).
\end{align}
\end{proof}

\begin{corollary}
    If $\operatorname{rank}(\bC) = N$, $\operatorname{rank}(\Delta) = N-1$. 
\end{corollary}

\begin{remark}
While $\Delta$ is a positive semi-definite matrix, it is typically low-rank when $N$ is large. The eigenvalues of $\Delta$, denoted by $\alpha_k$, satisfy the ordering:
\begin{equation}
\alpha_1 \geq \dots \geq \alpha_{\operatorname{rank}(\bC)-1} > 0,
\end{equation}
with $\alpha_k = 0$ for all $k \geq \operatorname{rank}(\bC)+1$.  In addition, running the RBD method for $N-1$ iterations is sufficient to accurately capture all eigenvalues of $\Delta$. 
\end{remark}
Finally, we analyze the computational and storage cost.  The sparse matrix $\bE$ is constructed in $O(N^2)$ flops. The RBD process on $\bE$, performed over $d$ iterations, requires $O(N^2d^2)$ for MGS and $O(N^2d^2  + Nd^3)$ for error calculation. Since $d \le N-1$, the entire cost for RBD is $O(N^2d^2)$. The construction of $\widetilde{\bC}$ and its SVD require $O(Nd^2)$ flops. Thus, the overall cost of our algorithm is $\mathcal{O}(N^2d^2)$. 

For the storage requirements, the sparse matrix $\bE \in \mathbb{R}^{\frac{N(N-1)}{2}\times N}$ contains $N(N-1)$ non-zero elements, resulting in a storage cost of $O(N^2)$ for its efficient representation. The algorithm must also store several other matrices: $\bC \in \mathbb{R}^{N \times N}$, $\bY \in \mathbb{R}^{\frac{N(N-1)}{2} \times d}$, $\bT \in \mathbb{R}^{d \times N}$, $\widetilde{\bC} \in \mathbb{R}^{d \times d}$, the singular values $\{\sigma_i\}_{i=1}^d$, and the matrix $\bY\bV \in \mathbb{R}^{\frac{N(N-1)}{2} \times d}$.Assuming $N$ is large, the storage requirement is dominated by the matrices $\bY$ and $\bY\bV$. Therefore, the total storage cost of the proposed algorithm is $O(N^2d)$.

\subsection{Implementing the method}
Our workflow to obtain the approximate scheduling distribution $\widetilde{\mathcal{S}}$ directly from $C$ is shown in Algorithm \ref{alg:rbd_scheduling}. This method avoids constructing $\Delta$. An R routine to construct the scheduling distribution using the RBD method can be found at \url{https://github.com/HiddenHarmsHub/RBD-for-CJ-Design}. 

\begin{algorithm}[ht!]
\caption{Construction of approximate scheduling distribution $\widetilde{\mathcal{S}}$ using RBD}
\label{alg:rbd_scheduling}
\begin{algorithmic}[1]
\Require Objects $\{\lambda_i\}_{i=1}^N$ 

\Ensure Approximate scheduling distribution $\widetilde{\mathcal{S}}$ over all possible pairings

\State Construct a prior mean vector $\boldsymbol{\mu}$ and prior covariance matrix $C$, and place a distribution on $\blambda \sim N(\boldsymbol{\mu}, C)$.

\State Construct the reduced basis decomposition of $\Delta$ from $C$ using Algorithm \ref{alg:rb_spectral} to obtain the approximate eigenpair $\{\sigma_i, \bY \bV(:,i)\}_{i=1}^d$. 

\State Compute the approximate design probabilities $\tilde{q}_{ij} = \frac{\sum_c \left(\mathbf{yv}^{(c)}_{r}\right)^2 \sigma^{(c)}}{\sum_d \sigma^{(d)}}$. 

\end{algorithmic}
\end{algorithm}

\section{Empirical results} \label{sec: simulation study}
In this section, the standard and approximate algorithms (Algorithms \ref{alg:brute_force} and \ref{alg:rbd_scheduling} respectively) are used to generate the standard and approximate scheduling distributions under a range of conditions. We evaluate the speed and accuracy of the RBD based method compared to the standard method. 

\subsection{Scalability in size of study}
The computational viability of experimental designs for the Bradley--Terry model is heavily dependent on the number of objects, $N$, as the size of the required covariance matrix $\Delta$ grows quadratically with $N$. To demonstrate the scalability of our proposed Reduced Basis Decomposition framework, we perform a comparative simulation study against the baseline method of computing $\Delta$ and then performing SVD. 

We simulate scenarios with increasing $N$, ranging from small-scale studies ($N=8$) to high-dimensional cases ($N=128$) where standard decomposition techniques typically become intractable. To investigate the use of the method for different kinds of prior covariance structures, we simulate different covariance matrices. The first is a Graph Laplacian covariance structure, where we simulate an Erd\H{o}s--R\'enyi graph with $N$ nodes and the probability of an edge being placed between two nodes $p = 0.5$, i.e. $G(N, 0.5)$. We used $p = 0.5$ to ensure the resulting graphs are sufficiently connected, while retaining enough sparsity to be non-trivial. The Graph Laplacian matrix is given by $C=(D-A+I)^{-1}$, where $A$ is the adjacency matrix. This enforced smoothness between the study objects and can be used in Gaussian Markov Random Fields. The second is a Toeplitz covariance matrix, where $C_{ij} = 0.5^{|i-j|}$, which can be used where there is a prior autoregressive or ordering assumption. The final covariance structure is a Inverse-Wishart distribution with the identity matrix $I$ as the scale matrix, which represents studies where there is weakly informative prior information. For structure, we normalise the resulting matrix by $C' = D^{-\frac{1}{2}}CD^{-\frac{1}{2}}$, where $D$ is a diagonal matrix containing the elements on the diagonal of $C$. This allows us to provide a more fair comparison between simulations, as the variance of each objects' quality is the same. 

For each value of $N$, we generate 100 covariance matrices of each structure and calculate the scheduling distribution ${\mathcal{S}}$ using the standard method in Algorithm \ref{alg:brute_force} and the approximate scheduling distribution $\widetilde{\mathcal{S}}$ using the RBD method in Algorithm \ref{alg:rbd_scheduling}. In Table \ref{tab:performance}, we show the computation time (in seconds) for both the standard and RBD methods. We see considerable efficiency gains for the RBD method across of study sizes and covariance structures. The efficiency gains increase with the study size, where studies with 128 objects, we see a considerable improvement for the RBD method, which takes around 2 seconds compared to over 4 minutes with the standard method. For studies with 64 objects and under, in even the slowest simulations, it takes under a second to compute the approximate scheduling distribution with RBD, compared to over 10 seconds in the best performing simulations for the standard method. We also see that the time taken for the RBD method is consistent, with little variation in the time taken for specific values of $N$ or covariance structure. 

Fitting a log-log linear model to the time taken and the $N$, suggests that the time taken for the RBD method depends on $N^{2.81}$, which is similar to the theoretical results derived in the previous section. For the standard method, the time taken depends on $N^{4.24}$. We also generate 100 networks from a $G(512, 0.5)$ model and construct the correspondent covariance matrices using the graph Laplacian approach. Using the RBD method the mean time taken to construct the approximate scheduling distribution is 23 minutes. We are unable to construct the true scheduling distribution using the standard method, but predict it will take 11.5 days. 


\begin{table}[ht]
\centering
\caption{Performance Comparison Across Covariance Structures in seconds}\label{tab:performance}
\begin{tabular}{lrcccc}
\toprule
 &  & \multicolumn{2}{c}{RBD} & \multicolumn{2}{c}{Standard}\\
\cmidrule(lr){3-4} \cmidrule(lr){5-6}
Structure & N & Median & (Min, Max) & Median & (Min, Max)\\
\midrule
Laplacian & 8 & 0.0008 & (0.0007, 0.0015) & 0.0025 & (0.0023, 0.0106)\\
 & 16 & 0.0023 & (0.0021, 0.0088) & 0.0378 & (0.035, 0.0583)\\
 & 32 & 0.0100 & (0.0095, 0.0155) & 0.660 & (0.637, 1.17)\\
 & 64 & 0.11 & (0.0995, 0.578) & 11.3 & (10.6, 12.0)\\
 & 128 & 2.17 & (1.60, 3.19) & 251 & (227, 1080)\\
\midrule
Toeplitz & 8 & 0.0008 & (0.0007, 0.0012) & 0.0027 & (0.0024, 0.0149)\\
 & 16 & 0.0025 & (0.0023, 0.0086) & 0.0376 & (0.0337, 0.0431)\\
 & 32 & 0.0077 & (0.0075, 0.0115) & 0.533 & (0.515, 0.834)\\
 & 64 & 0.112 & (0.102, 0.609) & 11.5 & (11.0, 11.9)\\
 & 128 & 2.46 & (1.79, 3.41) & 276 & (247, 317)\\
\midrule
Inverse Wishart & 8 & 0.0009 & (0.0008, 0.00160) & 0.0027 & (0.0025, 0.0132)\\
 & 16 & 0.0023 & (0.0022, 0.00250) & 0.0375 & (0.0338, 0.0417)\\
 & 32 & 0.0078 & (0.0075, 0.008) & 0.537 & (0.530, 0.856)\\
 & 64 & 0.111 & (0.0949, 0.578) & 11.7 & (10.8, 14.8)\\
 & 128 & 2.22 & (1.74, 3.04) & 418 & (254, 486)\\
\bottomrule
\end{tabular}
\end{table}

\begin{figure}[h!]
    \centering

    \begin{subfigure}{0.49\textwidth} 
        \centering
        \includegraphics[width=\textwidth]{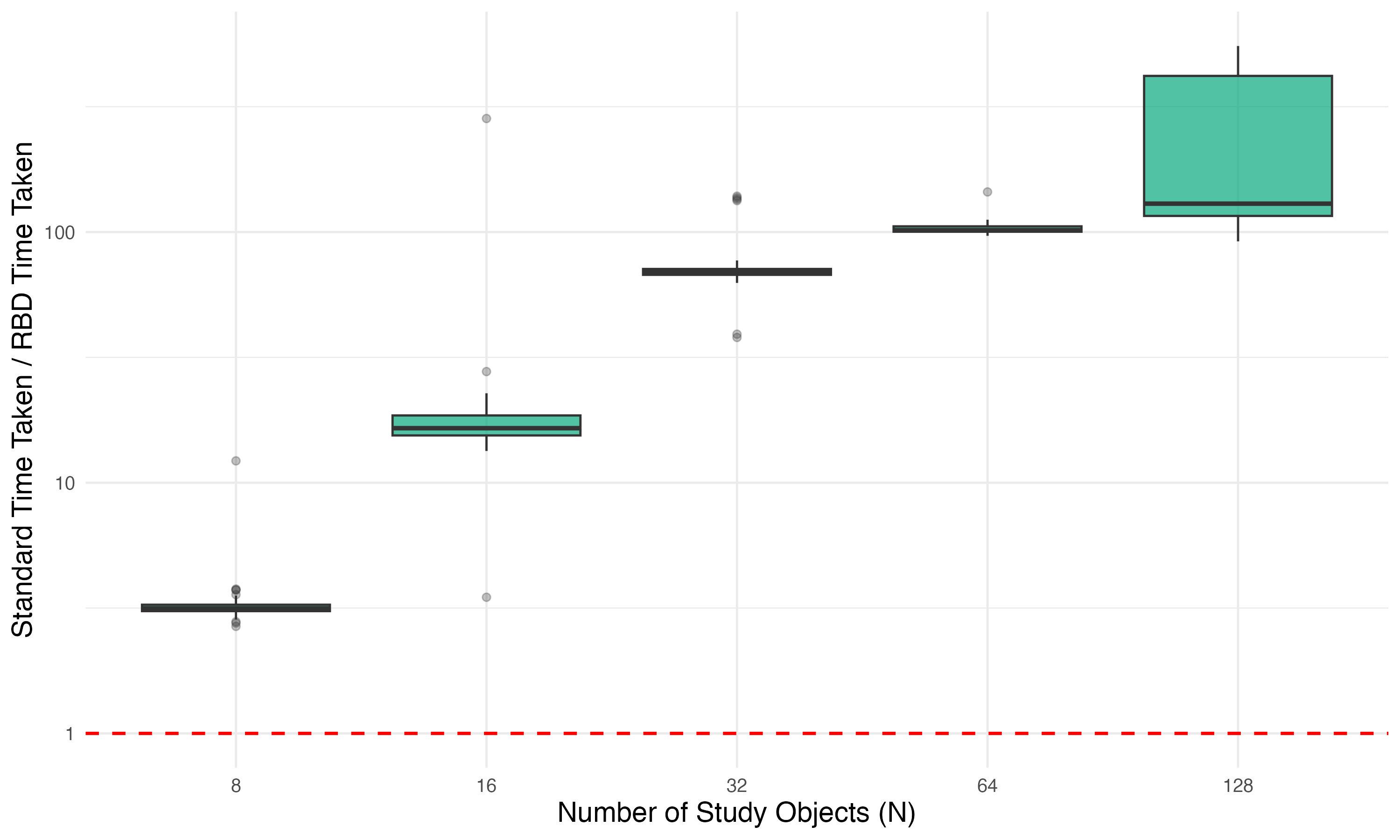}
        \caption{Laplacian Matrix Time Comparison}
        \label{fig:laplacian_speedup}
    \end{subfigure}
    \begin{subfigure}{0.49\textwidth}
        \centering
        \includegraphics[width=\textwidth]{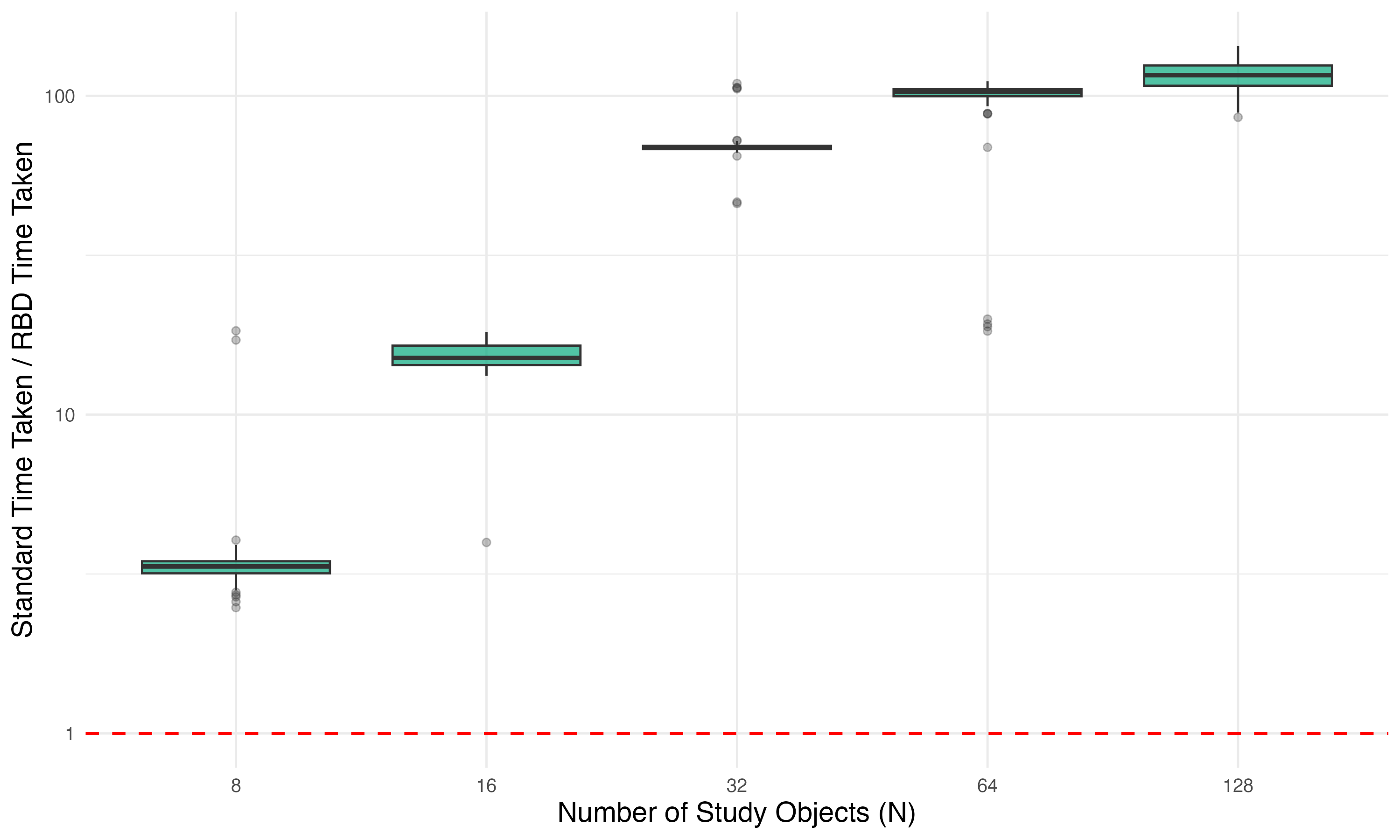}
        \caption{Toeplitz Matrix Time Comparison}
        \label{fig:toeplitz_speedup}
    \end{subfigure}

    \vspace{1em} 


    \begin{subfigure}{0.49\textwidth} 
        \centering
        \includegraphics[width=\textwidth]{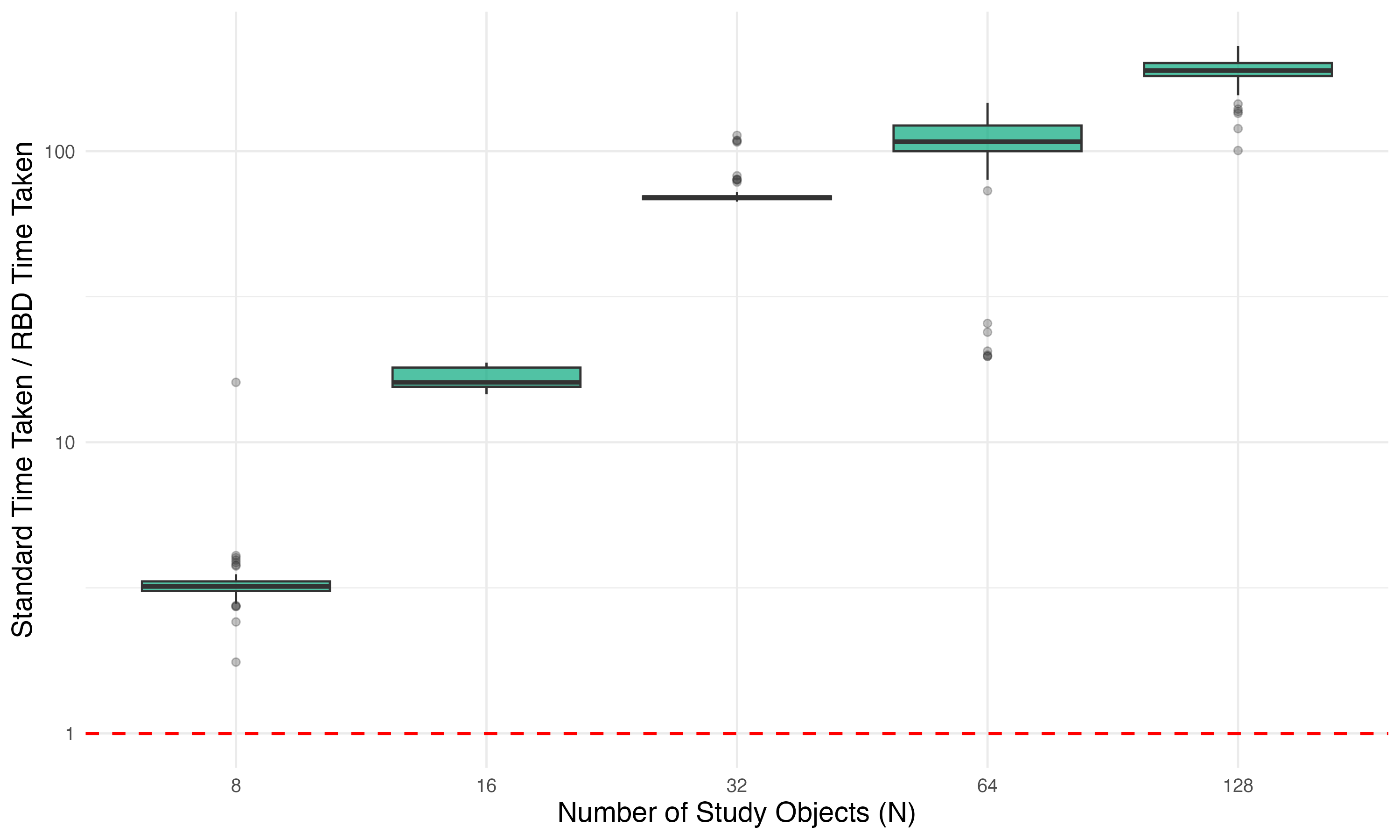}
        \caption{Inverse Wishart Distribution Time Comparison} 
        \label{fig:inverse_wishart_speedup}
    \end{subfigure}

    \caption{Comparison of speed up of the RBD method compared to the standard method for studies with different numbers of objects (N) in. The subfigures show (a) Laplacian, (b) Toeplitz, and (c) Inverse Wishart. Note the y-axes are on the log scale. }
    \label{fig:speedup}
\end{figure}

We also see a considerable improvement for the RBD method in Figure \ref{fig:speedup}, which shows the ratio of the time taken to compute the design probabilities using the standard to the RBD method. For all covariance structures, when there are 64 objects or more in the study, the RBD method is 100 times faster than the standard method.







\begin{figure}[h!]
    \centering

    \begin{subfigure}{0.49\textwidth} 
        \centering
        \includegraphics[width=\textwidth]{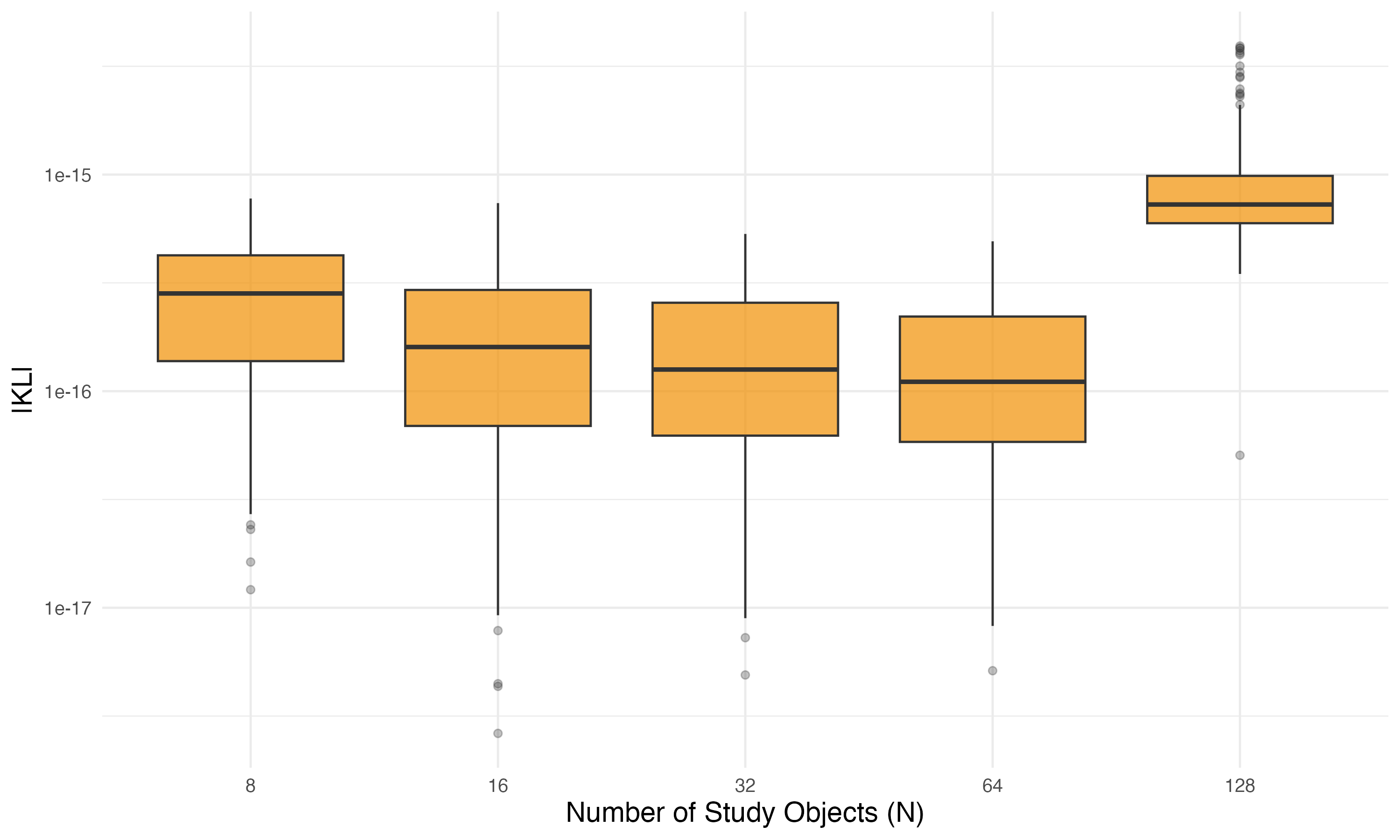}
        \caption{Laplacian Matrix Time Comparison}
        \label{fig:laplacian_kl}
    \end{subfigure}
    \begin{subfigure}{0.49\textwidth}
        \centering
        \includegraphics[width=\textwidth]{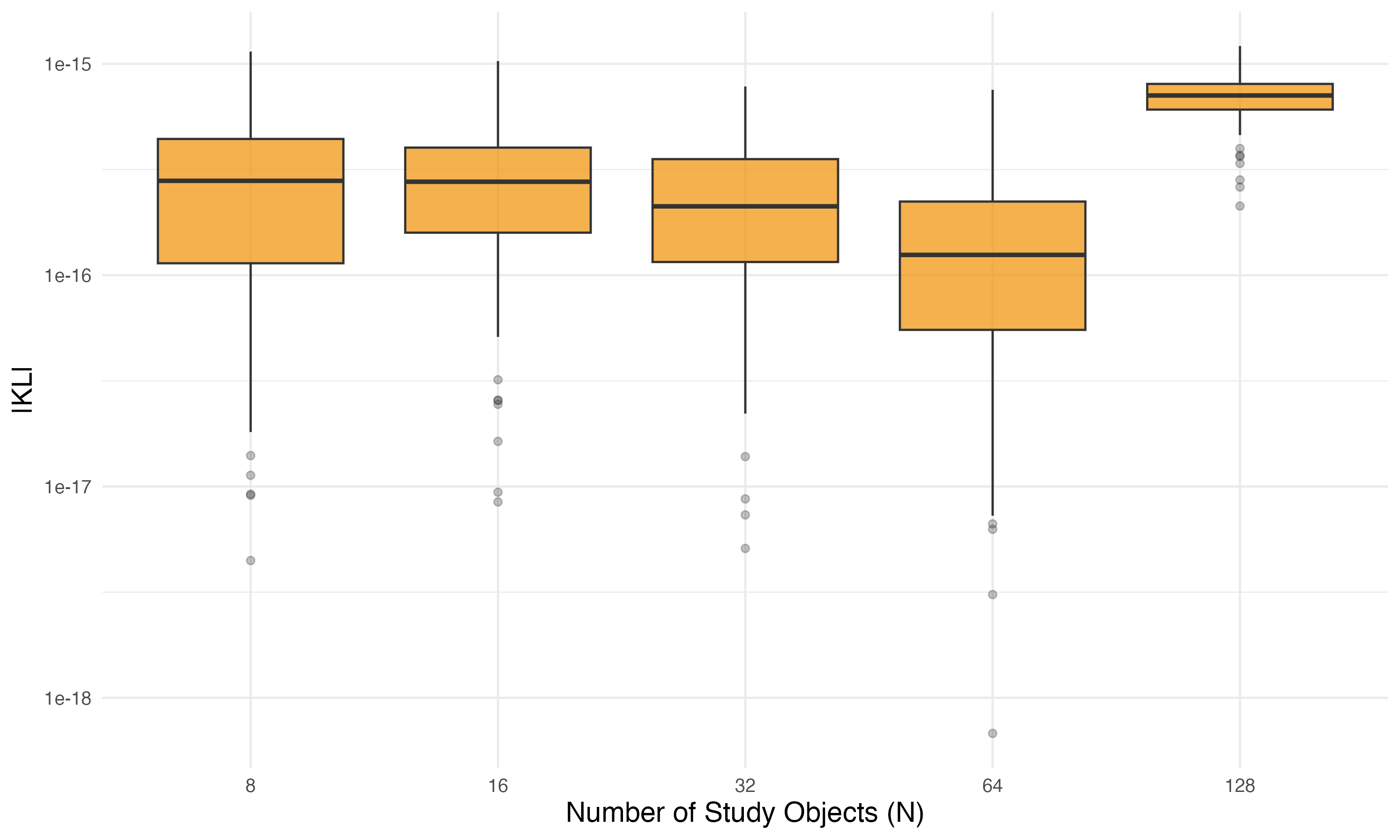}
        \caption{Toeplitz Matrix Time Comparison}
        \label{fig:toeplitz_kl}
    \end{subfigure}

    \vspace{1em} 


    \begin{subfigure}{0.49\textwidth} 
        \centering
        \includegraphics[width=\textwidth]{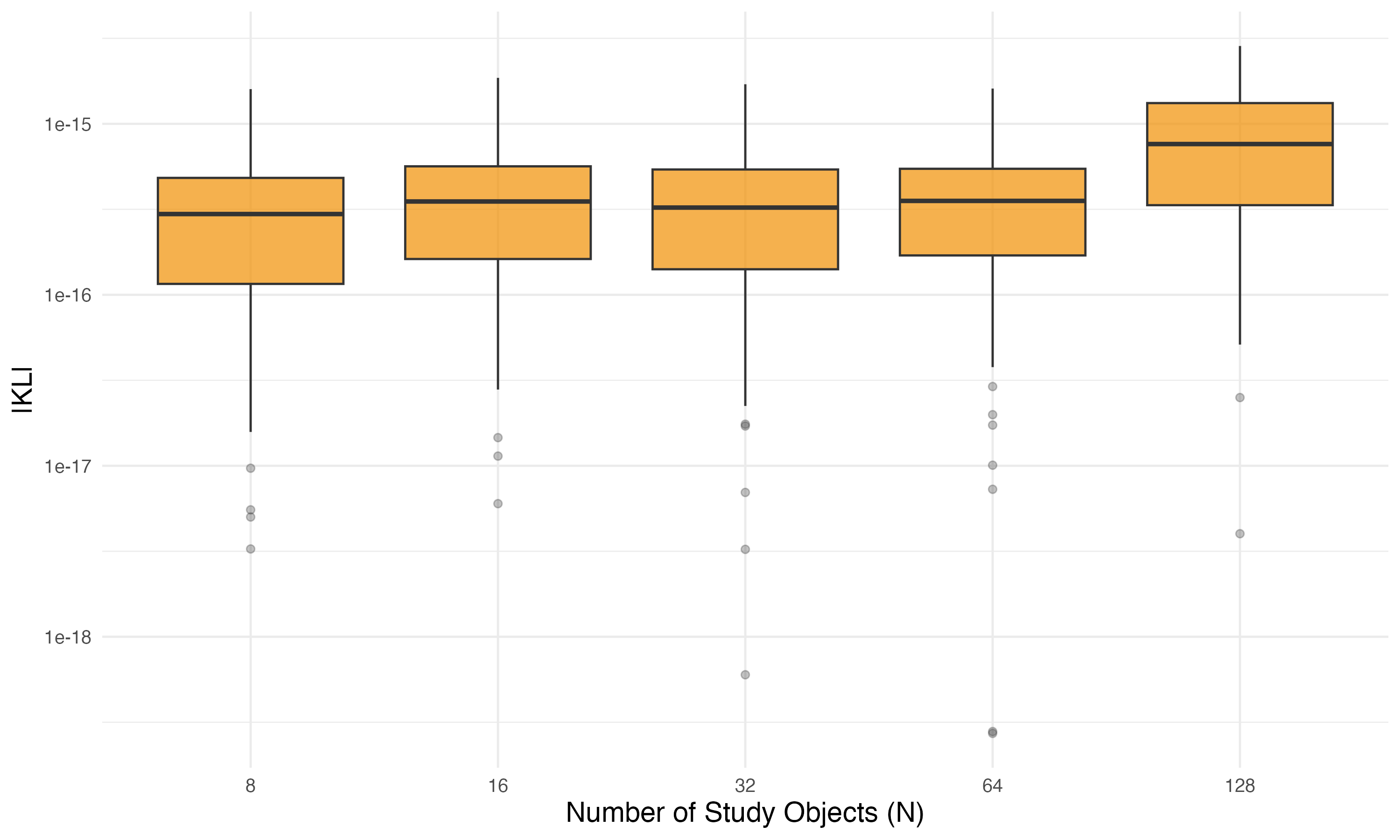}
        \caption{Inverse Wishart Distribution Time Comparison} 
        \label{fig:inverse_wishart_kl}
    \end{subfigure}

    \caption{Comparison of KL diveregnce for studies with different numbers of objects (N).  The subfigures show (a) Laplacian, (b) Toeplitz, and (c) Inverse Wishart. Note the y-axes are on the log scale. }
    \label{fig:kl}
\end{figure}

Finally, we compute the the Kullback-Leibler (KL) divergence for each method. For the scheduling distribution $\mathcal{S}$ computed via the standard method, and the approximate scheduling distribution $\widetilde{\mathcal{S}}$ computed via the RBD method, the KL divergence is given by
$$
D_{KL}({S} || \widetilde{S}) = \sum_{i=1}^N\sum_{j < i} S(i, j) \log \left( \frac{S(i, j)}{\widetilde{S}(i, j)} \right).
$$
This allows us to assess how well $\widetilde{\mathcal{S}}$ approximates the true distribution $\mathcal{S}$. In Figure \ref{fig:kl} we compute the KL divergence for the different study sizes and covariance structures. We find that there is negligible discrepancy between the true and approximating distributions, with the average KL divergence for all study sizes being between 1$\times 10^{-16}$ and 1$\times 10^{-15}$. There is no evidence of the covariance structure affecting the KL divergence. The KL divergence also remains the same for the majority of study sizes, except for the largest studies which have a slightly worse approximation, although this is still of the order 1$\times 10^{-15}$.

\subsection{Sensitivity to the sparsity of the prior covariance matrix}
To evaluate the method's performance under different connectivity regimes, we conducted a second set of simulations where the edge probability $p$ ranges from 0 to 1. This parameter acts as a proxy for the information density of the prior, scaling the complexity of the resulting covariance matrix $\Delta$ from a sparse, block-diagonal form to a fully dense matrix. For each value of $N$ and $p$ we generate 100 Erd\H{o}s--R\'enyi graphs and construct the covariance matrix by computing the regularised Graph Laplacian of the network $C=(D-A+I)^{-1}$, where $A$ is the adjacency matrix.

In Table \ref{tab:laplacian}, we report the mean computation time (in seconds), mean speedup (the computation time for standard method divided by the computation time for the RBD) and the mean Kullback-Leibler (KL) divergence for each method. For all values of sizes and sparsities of the covariance matrix, we see the KL divergence is negligible and at most $1.11\times 10^{-5}$. This suggests our method is not sensitive to the sparsity of the covariance matrix, and is suitable for complex prior distributions. We also see the time taken to compute the scheduling distribution does not vary as the level of sparsity changes. For the studies with 128 objects, the RBD method takes 2-3 seconds for all values of $p$, compared to the standard methods which takes 5-8 minutes. 

\begin{table}[ht]
\centering
\begin{tabular}{rrrrrr}
  \toprule
$N$ & $p$ & RBD (s) & Standard (s) & Speedup (Standard/RBD) & KL \\ 
  \midrule
  8 & 0.10 & $<0.01$ & $<0.01$ & 3.20 & $2.34\times 10^{-16}$ \\ 
    8 & 0.20 & $<0.01$ & 0.01 & 8.80 & $2.45\times 10^{-16}$ \\ 
    8 & 0.30 & $<0.01$ & $<0.01$ & 3.08 & $2.60\times 10^{-16}$ \\ 
    8 & 0.40 & $<0.01$ & 0.01 & 9.15 & $2.29\times 10^{-16}$ \\ 
    8 & 0.50 & $<0.01$ & $<0.01$ & 3.23 & $2.57\times 10^{-16}$ \\ 
    8 & 0.60 & $<0.01$ & $<0.01$ & 3.26 & $2.97\times 10^{-16}$ \\ 
    8 & 0.70 & $<0.01$ & $<0.01$ & 3.23 & $2.03\times 10^{-16}$ \\ 
    8 & 0.80 & $<0.01$ & $<0.01$ & 3.48 & $2.59\times 10^{-16}$ \\ 
   16 & 0.10 & $<0.01$ & 0.03 & 15.87 & $2.41\times 10^{-16}$ \\ 
   16 & 0.20 & $<0.01$ & 0.04 & 17.86 & $2.08\times 10^{-16}$ \\ 
   16 & 0.30 & $<0.01$ & 0.04 & 16.31 & $2.10\times 10^{-16}$ \\ 
   16 & 0.40 & $<0.01$ & 0.04 & 16.99 & $2.38\times 10^{-16}$ \\ 
   16 & 0.50 & $<0.01$ & 0.04 & 16.80 & $2.33\times 10^{-16}$ \\ 
   16 & 0.60 & $<0.01$ & 0.05 & 19.57 & $2.01\times 10^{-16}$ \\ 
   16 & 0.70 & $<0.01$ & 0.04 & 16.65 & $2.30\times 10^{-16}$ \\ 
   16 & 0.80 & $<0.01$ & 0.04 & 18.16 & $2.11\times 10^{-16}$ \\ 
   32 & 0.10 & 0.01 & 0.67 & 67.34 & $1.51\times 10^{-16}$ \\ 
   32 & 0.20 & 0.01 & 0.67 & 68.25 & $1.49\times 10^{-16}$ \\ 
   32 & 0.30 & 0.01 & 0.67 & 66.90 & $1.71\times 10^{-16}$ \\ 
   32 & 0.40 & 0.01 & 0.71 & 71.52 & $1.54\times 10^{-16}$ \\ 
   32 & 0.50 & 0.01 & 0.68 & 67.65 & $1.43\times 10^{-16}$ \\ 
   32 & 0.60 & 0.01 & 0.71 & 71.42 & $1.60\times 10^{-16}$ \\ 
   32 & 0.70 & 0.01 & 0.68 & 67.58 & $1.55\times 10^{-16}$ \\ 
   32 & 0.80 & 0.01 & 0.57 & 67.98 & $1.50\times 10^{-16}$ \\ 
   64 & 0.10 & 0.14 & 14.69 & 104.74 & $1.54\times 10^{-16}$ \\ 
   64 & 0.20 & 0.13 & 12.20 & 102.29 & $1.43\times 10^{-16}$ \\ 
   64 & 0.30 & 0.16 & 14.49 & 102.26 & $1.51\times 10^{-16}$ \\ 
   64 & 0.40 & 0.13 & 11.38 & 98.03 & $1.19\times 10^{-16}$ \\ 
   64 & 0.50 & 0.12 & 11.29 & 101.03 & $1.33\times 10^{-16}$ \\ 
   64 & 0.60 & 0.11 & 11.29 & 103.34 & $1.47\times 10^{-16}$ \\ 
   64 & 0.70 & 0.10 & 9.95 & 106.53 & $1.33\times 10^{-16}$ \\ 
   64 & 0.80 & 0.10 & 10.06 & 107.71 & $1.37\times 10^{-16}$ \\ 
  128 & 0.10 & 2.55 & 517.80 & 204.61 & $1.11\times 10^{-15}$ \\ 
  128 & 0.20 & 2.99 & 443.32 & 148.54 & $7.86\times 10^{-16}$ \\ 
  128 & 0.30 & 2.91 & 512.59 & 179.46 & $9.40\times 10^{-16}$ \\ 
  128 & 0.40 & 2.09 & 376.96 & 182.15 & $9.17\times 10^{-16}$ \\ 
  128 & 0.50 & 2.18 & 340.81 & 159.36 & $7.97\times 10^{-16}$ \\ 
  128 & 0.60 & 2.04 & 459.24 & 226.20 & $1.12\times 10^{-15}$ \\ 
  128 & 0.70 & 2.22 & 402.67 & 186.81 & $9.99\times 10^{-16}$ \\ 
  128 & 0.80 & 2.55 & 389.66 & 154.52 & $8.74\times 10^{-16}$ \\ 
   \bottomrule
\end{tabular}
\caption{The sensitivity of the standard and RBD methods for generating design probabilities for Graph Laplacian covariance matrices.} 
\label{tab:laplacian}
\end{table}

\subsection{Sensitivity to the tolerance parameter}
\begin{figure}
        \centering
        \includegraphics[width = 0.63\textwidth]{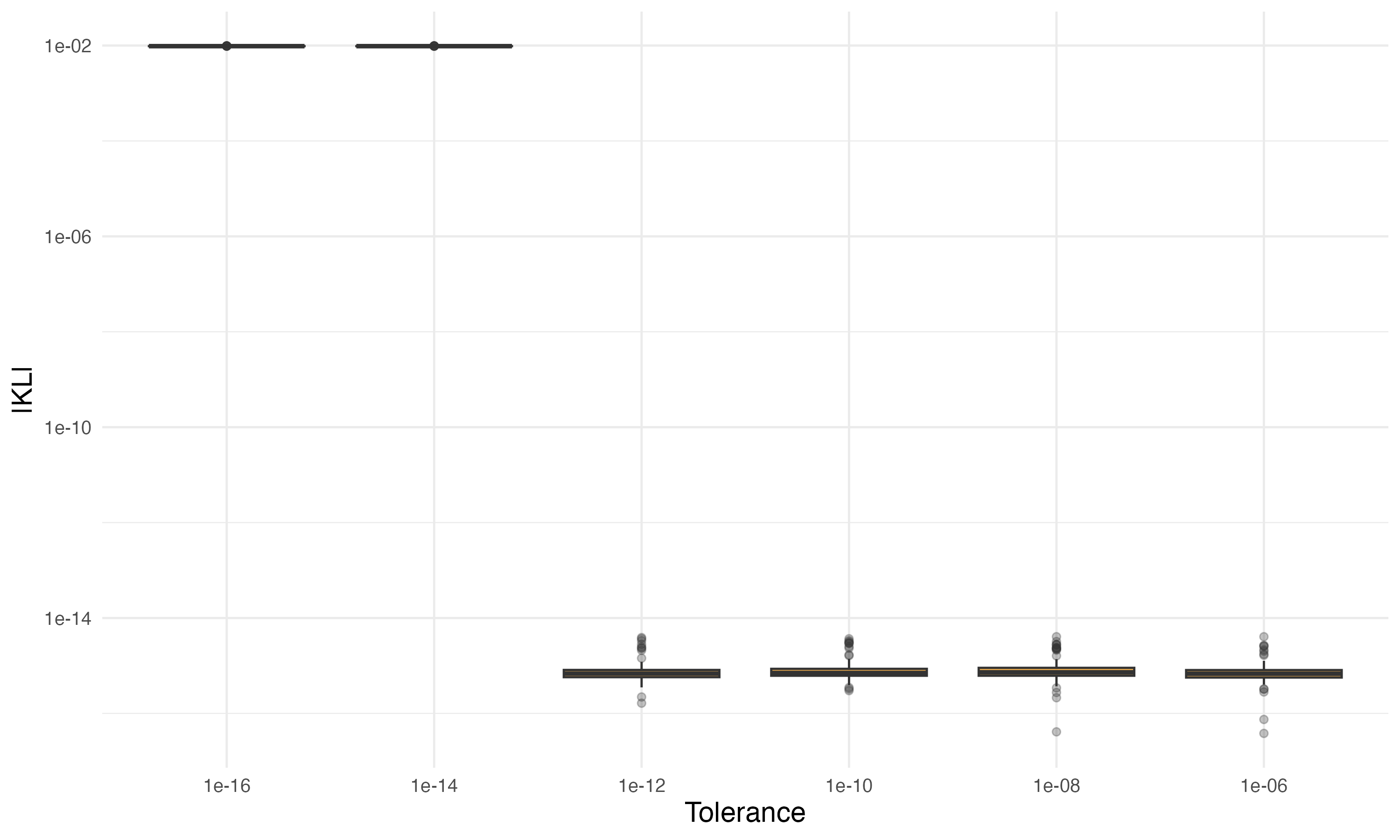}
        \caption{Comparisons of the KL divergence for different values of the tolerance parameter $\varepsilon_R$.} 
        \label{fig: tol comparison}
\end{figure}

We conduct a sensitivity analysis on the tolerance parameter, $\varepsilon_R$, used within the RBD greedy selection step. This threshold determines the sparsity of the reduced basis and, consequently, the dimension of the auxiliary eigenvalue problem. Although typically $\varepsilon_R$ is set to be 1$\times 10^{-6}$, we systematically vary $\varepsilon_R$ to evaluate how the algorithm behaves under different approximation constraints. For each of $\varepsilon_R = \{1\times 10^{-6}, 1\times 10^{-8}, 1\times 10^{-10}, 1\times 10^{-12}, 1\times 10^{-14}, 1\times 10^{-16}\}$, we generate 100 random graphs with 128 nodes and edge probability $p = 0.5$ and construct the corresponding covariance matrices based on the graph Laplacian. Figure \ref{fig: tol comparison} shows the relationship between the chosen tolerance and the KL divergence. The method is robust to the choice of $\varepsilon_R$ between 1$\times 10^{-6}$ and 1$\times 10^{-12}$, with the KL divergence being 3$\times 10^{-16}$ regardless of the choice of the tolerance parameter. We see that the method does not work for $\varepsilon_R = \{1\times 10^{-14}, 1\times 10^{-16}\}$ due to precision errors in R. 

\section{Applications to real data} \label{sec: application}
We now demonstrate our RBD method on two real data sets. This allows us to evidence the scalability and speed of the method in practical situations. 

\subsection{Scheduling for large spatial studies}
To demonstrate the utility of the proposed Reduced Basis Decomposition method in a real-world setting, we consider the large-scale comparative judgement study presented by \cite{Seymour2022}. The objective of the study was to use comparative judgement to quantify levels of urban deprivation across Dar es Salaam, Tanzania, overcoming the limitations of traditional household surveys which are often costly, slow, and prone to rapid obsolescence in fast-growing cities. In the study, spatial information was used to construct the prior covariance matrix. 

The study employed a pairwise comparison approach to elicit local knowledge from citizens. The objects of interest were $N=452$ administrative sub-wards within the city. Judges were presented with pairs of sub-wards (displayed via imagery or identified by name) and asked to select the more affluent of the two. \cite{Seymour2022} collected approximately 75,000 comparisons to achieve stable estimates, utilizing a spatial prior distribution to borrow strength between neighbouring regions. A scheduling algorithm could have reduced the number of comparisons that were collected, but it was not possible to construct without our RBD method, especially in the field in Tanzania. 

This application is particularly relevant to our proposed method due to the dimensionality of the parameter space. With $N=452$ objects, the full design space consists of $\begin{pmatrix}
    452 \\2 
\end{pmatrix} = 101,296$ possible pairs, meaning there are over 5 billion pairs of pairs. Constructing a static design for this study has not previously been possible due to the size of this matrix. 

We follow \cite{Seymour2022} and construct an adjacency matrix $A$ from the sub-wards of Dar es Salaam by treating the wards as nodes and placing edges between geographically adjacent wards. The prior distribution on the deprivation parameters of each ward is given by $\boldsymbol{\lambda} \sim N(\boldsymbol{0}, \, C)$, where $C = D^{-\frac{1}{2}}\Lambda D^{-\frac{1}{2}}$, $\Lambda = e^A$,  $A$ is the network's adjacency matrix, and $D$ is a diagonal matrix containing the elements on the diagonal of $\Lambda$. Using the matrix exponential of the adjacency matrix as the basis of the covariance matrix assigns higher prior covariance to pairs of subwards that are well-connected in the network, and low prior covariance to pairs that are poorly connected. 

We use the RBD algorithm shown in Algorithm \ref{alg:rbd} to generate the approximate scheduling distribution $\widetilde{\mathcal{S}}$ from the covariance matrix $C$. On a 2022 M1 iMac, this takes 6 minutes 25 seconds and avoids computing the pairwise covariance matrix $\Delta$, which has dimension 101,296 $\times$ 101,296.

\subsection{Scheduling for classroom based peer learning}
Comparative judgement is perhaps most used in educational contexts, as a form of peer learning and assessment. Under the theory of ``Learning by Evaluation'', students carry out an exercise and then compare pairs of answers \citep{Bartholomew2020}. The act of comparing peer work allows students to implicitly internalize quality standards and construct a robust understanding of the assessment criteria, often more effectively than through passive rubric analysis \citep{Hoffelinck2025}.

A common implementation of this Learning by Evaluation involves a synchronous, two-phase classroom activity. In the first phase, students perform a set number of comparisons on peer assignments (e.g., mathematical proofs, written texts, or explanations of concepts). This is followed by a class discussion on the topic and comparisons. Finally, students return for a second phase of judging to refine the grading and consolidate learning.

To maximize the statistical efficiency of the second phase, the design probabilities should ideally be updated based on the information gained in the first phase, specifically, by constructing the design from the posterior covariance matrix of the initial comparisons. However, in a live classroom setting, the break between phases where the discussion happens is typically short. Standard spectral decomposition methods are often too computationally intensive to update the design matrix for a large cohort in real-time.

We demonstrate the utility of our proposed method using data from \cite{JonesSirl2017}, where undergraduate mathematics students assessed peer responses to calculus conceptual problems. The data set consists of 3,258 comparisons of 139 students' written responses to a first year calculus assignment. The comparisons were made by the first year student who undertook the assignment as part of a Learning by Evaluation study, with the aim of improving both their understanding of calculus and ability to write formal mathematical arguments. There are around 23 comparisons for each assignment. 

We fit the Bradley-Terry model to the first 700 comparisons (approximately 20\% of the total number of comparisons), using $\lambda_i \sim N(0, 5^2), \, i = 1,\ldots, 139$ as the prior distributions on the quality of the assignments. We consider this as the first phase of the class, which is followed by an in-lecture activity or discussion. To run the second phase efficiently, we place a joint prior distribution on the quality parameters $\boldsymbol{\lambda} \sim N(\boldsymbol{0}, C)$, where $C$ is the posterior covariance matrix from the first phase. We can then construct the approximate scheduling distribution $\widetilde{\mathcal{S}}$ and show the students pairs of assignments according to this distribution. Reading in the data and fitting the Bradley--Terry model takes 10.6 seconds. Using the standard method, it takes 15 minutes 43 seconds to construct $B$ and then compute $\mathcal{S}$, for a total workflow of 15 minutes 54 seconds. Using our RBD method, it takes 4.08 seconds to construct $\widetilde{\mathcal{S}}$ directly from $C$, for a total workflow time of 14.68 seconds, thus reducing the time taken by 98\%. 

Consequently, the computational cost of the standard approach renders efficient data collection in the second phase impossible. With a lag of over fifteen minutes, the standard method effectively becomes unfeasible for live classes, as teachers cannot use up that much of the teaching time in the lesson. Our method makes it possible to carry out more efficient data in Learning by Evaluation settings. 

\section{Discussion} \label{sec: discussion}
In this paper, we have developed a scalable method for constructing optimal static designs in pairwise comparative judgement studies using reduced basis decomposition. Our approach circumvents the construction and decomposition of the high-dimensional pairs-of-pairs covariance matrix $\Delta$, achieving computational efficiency gains of two to three orders of magnitude compared to the standard method while maintaining negligible approximation error.

The empirical results demonstrate that our method scales effectively with study size, with computational complexity of $O(N^3 d)$ compared to $O(N^6)$ for standard spectral decomposition. For studies with 128 objects, the RBD method completes in approximately 2 seconds compared to over 4 minutes for the standard approach, a speedup factor exceeding 100. Importantly, this efficiency does not come at the cost of accuracy: the Kullback-Leibler divergence between the approximate and exact scheduling distributions remains below $10^{-15}$ across all simulation scenarios, indicating that the approximation is effectively exact to machine precision. The method also exhibits robustness to the choice of tolerance parameter $\varepsilon_R$ across a wide range of values ($10^{-6}$ to $10^{-12}$), simplifying practical implementation.

The practical applications highlight the method's utility in real-world scenarios. In the Dar es Salaam deprivation study, our approach enabled design construction for 452 sub-wards, involving over 5 billion pairs of pairs, in under 7 minutes, where standard methods would be computationally infeasible. In the classroom peer learning example, reducing workflow time from over 15 minutes to approximately 15 seconds makes adaptive two-phase designs pedagogically feasible, allowing teachers to update designs during brief class breaks without disrupting instruction flow.

A potential benefit of static scheduling designs in peer learning contexts, is that unlike most other comparative judgement scenarios, we care not only about the statistical efficiency of the design but also about the pedagogical experience of the student acting as judge. Comparisons between very similar pieces of work may be statistically informative but offer limited learning value, as the student has little to reflect on when the differences are negligible. Conversely, comparisons between work of vastly different quality may be too obvious to prompt meaningful engagement with the assessment criteria. A well-constructed scheduling distribution can balance these considerations.

Our theoretical contributions establish tight bounds on eigenvalue approximations (Theorem~3.1) and characterize the relationship between $\operatorname{rank}(\Delta)$ and $\operatorname{rank}(\bC)$ (Theorem~3.2), providing rigorous guarantees on approximation quality. This analysis also reveals that $\operatorname{rank}(\Delta) \le \operatorname{rank}(\bC)$, implying that $\Delta$ is typically low-rank for large $N$, particularly when $\bC$ is itself low-rank. Notably, even when $\bC$ is full rank, $\operatorname{rank}(\Delta)$ is capped at $N-1$. Consequently, $N-1$ iterations of the RBD algorithm are sufficient to capture the full spectrum of $\Delta$. Furthermore, these theoretical results provide a principled upper bound on the required number of iterations for practitioners who choose to employ truncated PCA or SVD.

A limitation of our approach is that it assumes a static design framework with known prior information. While this accommodates sequential multi-phase designs where interim posteriors inform subsequent scheduling, it does not address fully online adaptive designs that update after every comparison. However, as noted in Section~1, static designs avoid the reliability inflation and increasing comparison difficulty problems associated with adaptive comparative judgement \citep{Bramley2018,Jones2023}. Another consideration is numerical precision: the sensitivity analysis (Section~4.3) demonstrates that very small tolerance values ($\varepsilon_R < 10^{-12}$) can lead to precision errors in standard numerical computing environments, though the method remains robust across practically relevant tolerance ranges. Finally, regarding computational complexity, if $\operatorname{rank}(\bC) \ll N$, full precision for the few non-zero eigenvalues of $\Delta$ still requires $N-1$ iterations of the RBD method. Using fewer iterations will yield approximated spectral results rather than the exact values. 

A broader limitation concerns the optimality of the underlying design criterion itself. Although D-optimal designs can now be computed in a classical framework \citep{Roettger2025}, we are not aware of any method to contract Bayesian D-optimal designs. The computational efficiency of our RBD method may allow for Bayesian D-optimal designs to be constructed for the first time.  The principal-component-based design of \citet{Seymour25} assigns higher scheduling probability to pairs exhibiting greater prior variance, but it remains an open question whether this criterion is optimal in any formal sense, such as D-optimality or A-optimality. Establishing optimality would require comparing alternative design criteria, yet such comparisons have been hindered by the computational expense of constructing designs under the standard approach. By reducing this computational burden, our RBD method opens the door to systematic investigation of design optimality for comparative judgement studies. Researchers can now feasibly construct and evaluate competing design criteria for moderately large studies, enabling future work to establish theoretical foundations for optimal experimental design in the Bradley-Terry framework with correlated priors.


Our method removes a significant computational barrier to implementing optimal experimental designs in comparative judgement studies, enabling researchers and practitioners to leverage prior information and spatial or temporal structure in studies that were previously computationally intractable.

\section*{Acknowledgements}
This work was supported by a UKRI Future Leaders Fellowship [MR/X034992/1] and an EPSRC Mathematical Sciences Discipline Hopping Grant [UKRI2389]. We thank Cat Smith for her help in creating the coding workflow for this project. We thank Ian Jones for providing the Learning by Evaluation data. We would like to thank the reviewers for their insightful comments and constructive feedback, which have improved the quality of this work.

\bibliographystyle{apalike}
\bibliography{bibliography}

\end{document}